\documentclass[useAMS,usenatbib]{mnras}
\usepackage{journals}
\usepackage{graphicx}
\usepackage{color}
\usepackage{times}
\usepackage{hyperref}
\usepackage{amsmath}

\title[Supernovae and their host galaxies -- V]{Supernovae and their host galaxies -- V.
The vertical distribution of supernovae in disc galaxies}
\author[A.~A.~Hakobyan~et~al.]{A.~A.~Hakobyan,$^{1}$\thanks{E-mail:
hakobyan@bao.sci.am}
L.~V.~Barkhudaryan,$^{1}$
A.~G.~Karapetyan,$^{1}$
G.~A.~Mamon,$^{2}$
\newauthor
D.~Kunth,$^{2}$
V.~Adibekyan,$^{3}$
L.~S.~Aramyan,$^{1}$
A.~R.~Petrosian$^{1}$
and M.~Turatto$^{4}$
\\
$^{1}$Byurakan Astrophysical Observatory, 0213 Byurakan, Aragatsotn province, Armenia\\
$^{2}$Institut d'Astrophysique de Paris, Sorbonne Universit\'{e}s, UPMC Univ Paris 6 et CNRS, UMR 7095, 98 bis bd Arago, F-75014 Paris, France\\
$^{3}$Instituto de Astrof\'{i}sica e Ci\^{e}ncia do Espa\c{c}o, Universidade do Porto, CAUP, Rua das Estrelas, P-4150-762 Porto, Portugal\\
$^{4}$INAF -- Osservatorio Astronomico di Padova, Vicolo dell'Osservatorio 5, I-35122 Padova, Italy}
\begin{document}

\date{Accepted 2017 June 22. Received 2017 June 21; in original form 2017 April 21}

\pagerange{\pageref{firstpage}--\pageref{lastpage}} \pubyear{2017}

\maketitle

\label{firstpage}

\begin{abstract}
  We present an analysis of the height distributions of the different types of supernovae (SNe)
  from the plane of their host galaxies.
  We use a well-defined sample of 102 nearby SNe appeared inside
  high-inclined ($i \geq 85^\circ$), morphologically non-disturbed
  S0--Sd host galaxies from the Sloan Digital Sky Survey.
  For the first time, we show that in all the subsamples of spirals,
  the vertical distribution of core-collapse (CC) SNe is about twice closer to the plane of host disc
  than the distribution of SNe Ia.
  In Sb--Sc hosts, the exponential scale height of CC SNe is consistent with those of
  the younger stellar population in the Milky Way (MW) thin disc, while the scale height
  of SNe Ia is consistent with those of the old population in the MW thick disc.
  We show that the ratio of scale lengths to scale heights of the
  distribution of CC SNe is consistent with those of the resolved young stars with
  ages from $\sim10$~Myr up to $\sim100$~Myr in nearby edge-on galaxies and
  the unresolved stellar population of extragalactic thin discs.
  The corresponding ratio for SNe Ia is consistent with the same ratios of the
  two populations of resolved stars
  with ages from a few 100~Myr up to a few Gyr and from a few Gyr up to $\sim10$~Gyr,
  as well as with the unresolved population of the thick disc.
  These results can be explained considering the age-scale height relation of
  the distribution of stellar population and the mean age difference between Type Ia and CC SNe progenitors.
\end{abstract}


\section{Introduction}
\label{intro}

\defcitealias{2012A&A...544A..81H}{I}

The detailed understanding of the spatial distribution of Supernovae (SNe)
in galaxies provides a strong possibility to find the links between the nature
of their progenitors and host stellar populations
\citep[e.g.][]{1997AJ....113..197V,2000ApJ...542..588I,
2005AJ....129.1369P,2008MNRAS.390.1527A,
2008A&A...488..523H,2009A&A...508.1259H,2012ApJ...759..107K,
2013Ap&SS.347..365N,2014A&A...572A..38G,2015A&A...580A.131T,2016MNRAS.459.3130A}.
Such studies allow to constrain the important physical parameters of
the different SN progenitors like their masses
\citep[e.g.][]{2012MNRAS.424.1372A,2017A&A...597A..92K},
ages \citep[e.g.][]{1996ApJ...473..707M,2008MNRAS.388L..74F},
and metallicities \citep[e.g.][]{2011ApJ...731L...4M,2016A&A...591A..48G}.
For a comprehensive review addressing these issues, the reader is referred
to \citet[][]{2015PASA...32...19A}.

According to the properties of SNe progenitors,
they are divided into two general categories:
core-collapse (CC) and Type Ia (thermonuclear) SNe.
CC SNe are the colossal explosions that mark the violent deaths of young massive stars
\citep[e.g.][]{2003LNP...598...21T,2009ARA&A..47...63S,
2011MNRAS.412.1522S},\footnote{{\footnotesize According to the
spectral features in visible light, CC SNe are classified into three basic
classes \citep[e.g.][]{1997ARA&A..35..309F}:
hydrogen lines are visible in the spectra of Type II SNe, but in Types Ib and Ic SNe;
helium lines are seen in the spectra of SNe Ib, but in SNe Ic.
Subclass IIn SNe are dominated by narrow emission lines,
while subclass IIb SNe have transitional spectra closer to SNe II at early times,
then evolving to SNe Ib.}}
while SNe Ia are the explosive end in the evolution of binary stars
in which one of the stars is an older white dwarf (WD) and the other star
can be anything from a giant star to a WD
\citep*[for a comprehensive review about thermonuclear SNe, see][]{2014ARA&A..52..107M}.
Type Ia SNe result from stars of different ages
\citep[from $\sim0.5$~Gyr up to $\sim10$~Gyr, see][]{2012PASA...29..447M},
with longer progenitors lifetime than the progenitors of CC SNe
\citep[from a few Myr up to $\sim0.2$~Gyr
when including the evolution of stars in close binary systems, see][]{2017A&A...601A..29Z}.

Usually, the spatial distribution of SNe in S0--Sm galaxies is
studied with the reasonable assumption that all
CC SNe and the vast majority of SNe Ia belong to the disc, rather than the bulge
population \cite[e.g.][]{1997AJ....113..197V,2000ApJ...542..588I,
2005AJ....129.1369P,2008Ap.....51...69H,2008MNRAS.390.1527A,2013Sci...340..170W}.
Moreover, the distributions of SNe in the disc are studied assuming that the disc is infinitely thin
\citep[e.g.][]{2009A&A...508.1259H,2013Sci...340..170W}.
The height distribution of SNe from the disc plane is mostly neglected when
studying the host galaxies with low inclinations (close to face-on orientation)
assuming that the exponential scale length of the radial distribution
is dozens of times larger in comparison with the exponential scale height of SNe
\citep[e.g.][]{1998ApJ...502..177H}.

Direct measurements of the heights of SNe and estimates of the scales of their
vertical distributions in host galaxies with high inclination (close to edge-on orientation)
were performed only in a small number of cases
\citep[][]{1997PhDT........11M,2012MsT..........1M,2016AstL...42..495P}.
Mainly due to the small number statistics of SNe and inhomogeneous data of
their host galaxies, the comparisons of vertical distributions of the different
types of SNe resulted in statistically insignificant differences.
Therefore, while the detailed study of the vertical distributions in
edge-on galaxies has allowed to constrain ages, masses and
other physical parameters of their components
\citep[e.g.][]{2005AJ....130.1574S,2006AJ....131..226Y,2014ApJ...787...24B},
the lack of analogous studies on the distribution of various SN types has prevented
the determination of their parent populations via the direct comparison with
the nearby extragalacric discs and the thick/thin discs of the Milky Way (MW) galaxy
\citep[e.g.][]{2001ApJ...553..184C,2003AJ....125.1958L,2008ApJ...673..864J}.

The purpose of this paper is to address these questions properly
through an investigation of the vertical distributions of
the main classes of SNe in a nearby sample of 102 SNe and
their well-defined edge-on S0--Sd host galaxies
from the Sloan Digital Sky Survey-III \citep[SDSS-III;][]{2015ApJS..219...12A}.

This is the fifth article of the series and the content is as follows.
Sample selection and reduction are introduced in Section~\ref{sample}.
Section~\ref{discmodel} describes the stellar disc model that we use to fit our data.
All the results are discussed in Section~\ref{resdiscus}.
Section~\ref{concl} summarizes our conclusions.
To conform to values used in our data base \citep[][hereafter Paper~I]{2012A&A...544A..81H},
a cosmological model with $\Omega_{\rm m}=0.27$, $\Omega_{\rm \Lambda}=0.73$,
and $H_0=73 \,\rm km \,s^{-1} \,Mpc^{-1}$ Hubble constant \citep{2007ApJS..170..377S}
are adopted in this article.

\section{Sample selection and reduction}
\label{sample}

\defcitealias{2014MNRAS.444.2428H}{II}

In this paper, we composed our sample by cross-matching the coordinates of
classified Ia, Ibc\footnote{{\footnotesize `Stripped-envelope'
SNe of Types Ib and Ic, including the mixed Ib/c with uncertain subclassification,
are denoted as SNe Ibc.}},
and II SNe from the
Asiago Supernova Catalogue\footnote{{\footnotesize We use
the updated version of the catalogue, which includes all classified SNe exploded
before 2015 January 1.}}
\citep[ASC;][]{1999A&AS..139..531B}
with the footprint of SDSS Data Release 12 \citep[DR12;][]{2015ApJS..219...12A}.
All SNe are required to have equatorial coordinates.
We use SDSS DR12 and the approaches presented in Paper~\citetalias{2012A&A...544A..81H} to
identify the host galaxies and classify their morphological types.
It is worth noting that morphological classification of nearly edge-on galaxies
is largely based on the visible size of bulge relative to the disc
because other morphological properties, such as the shape of spiral arms
or presence of the bar, are generally obscured or invisible.
The morphologies of galaxies are restricted to S0--Sd types,
since we are interested in studying the vertical distribution of
SNe in host stellar discs.
A small number of Sdm--Sm host galaxies are not selected,
because they show no clear discs.

From the signs of galaxy--galaxy interactions,
we classify the morphological disturbances of the hosts
in the SDSS DR12 following the techniques described in detail
in \citet[][hereafter Paper~II]{2014MNRAS.444.2428H}.
We then exclude from this analysis any galaxy disc exhibiting
strong disturbances: interacting, merging, and post-merging/remnant.

Using the techniques presented in Paper~\citetalias{2012A&A...544A..81H},
we measure the apparent magnitudes and the geometry of host
galaxies.\footnote{{\footnotesize Instead using the data from
Paper~\citetalias{2012A&A...544A..81H}, which is based on the SDSS DR8,
for homogeneity we re/measure the magnitudes and the geometry of all host galaxies,
with additional new SN hosts included, based only on DR12.}}
In the SDSS $g$-band, we first construct isophotes, and then centred at
the each galaxy centroid position an elliptical aperture visually fitted to
the $25~{\rm mag~arcsec^{-2}}$ isophote.
We measure the apparent magnitudes, major axes ($D_{25}$), position angles (PA) of the major axes,
and elongations ($a/b$) of galaxies using these apertures.
In this analysis, we correct the magnitudes and $D_{25}$ for
Galactic and host galaxy internal extinction (see Paper~\citetalias{2012A&A...544A..81H}).

\subsection{Inclination}
\label{inclin}

\defcitealias{2016MNRAS.456.2848H}{III}

The main difficulty in measuring the vertical distribution of SNe above
the host stellar discs is that we have no way of knowing where along
the line of sight the SNe lie. This means that reliable measurements can only be done
in discs which are highly inclined, i.e., closer to an edge-on
orientation (e.g. $85^\circ \leqslant i \leqslant 90^\circ$).
In contrast to galaxies with lower inclination, the matter is complicated
by the difficulty of making an accurate determination of the inclination angle.
For these galaxies, the inclination cannot be measured
simply from the major and minor axes because the presence of a central bulge places
a limit on the axis-ratio even for a perfectly edge-on galaxy.

This problem with the bulge has been solved by using the axial ratio of
the exponential disc fits in the $g$-band provided by the SDSS
(from the model with $r^{1/4}$ bulge and exponential disc), i.e., \texttt{expAB\_g}.
Indeed, real stellar discs are not flat with negligible thicknesses,
but have some intrinsic width, and a proper measurement of the
inclination depends on this intrinsic ratio of the vertical and horizontal
axes of the disc, known as $q$.
Therefore, we calculate the inclinations of SNe host galaxies following
the formula
\begin{eqnarray}
\cos^2 i= \frac {(\texttt{expAB\_g})^2-q^2}{1-q^2} \,\ ,
\label{galincl}
\end{eqnarray}
where $i$ is the inclination angle in degrees
between the polar axis and the line of sight and $q$ is the intrinsic
axis-ratio of galaxies viewed edge-on.
According to \cite{1997A&AS..124..109P},
\begin{eqnarray}
q={\rm dex}[-(0.43+0.053\, t)]
\label{galincl_2}
\end{eqnarray}
for $-1\leq t\leq 7$, where $t$ is the morphological type code.
Using equations~(\ref{galincl}) and (\ref{galincl_2}), we restrict the inclinations
of host galaxies to $85^\circ \leqslant i \leqslant 90^\circ$.

All the selected SNe host galaxies are visually inspected because
sometimes bright stars projected nearby,
strong dust layers, bright nuclear/bulge emission, large angular sizes, etc.
do not allow the SDSS automatic algorithm to correctly determine the parameters of galaxies,
in particular the axis-ratio \texttt{expAB\_g}.
The host discs with a clearly seen dust layer, or without signs of non-edge-on spiral arms,
are selected as true edge-on galaxies. In other words, we exclude the discs whose
galactic plane is not aligned along the major axis of their fitted elliptical apertures
\citep[e.g. warped edge-on discs, see][]{2016MNRAS.461.4233R}.
As a result, we select 106 SNe in edge-on host galaxies.

\begin{table}
  \centering
  \begin{minipage}{79mm}
  \caption{Numbers of SNe as a function of morphological types of edge-on S0--Sd host galaxies.}
  \tabcolsep 5pt
  \label{table_SN_morph}
  \begin{tabular}{lrrrrrrrrrr}
  \hline
   &\multicolumn{1}{c}{S0}&\multicolumn{1}{c}{S0/a}&\multicolumn{1}{c}{Sa}
  &\multicolumn{1}{c}{Sab}&\multicolumn{1}{c}{Sb}&\multicolumn{1}{c}{Sbc}&\multicolumn{1}{c}{Sc}&\multicolumn{1}{c}{Scd}
  &\multicolumn{1}{c}{Sd}&\multicolumn{1}{r}{All}\\
  \hline
  Ia & 6 & 3 & 2 & 5 & 9 & 5 & 16 & 4 & 3 & 53 \\
  Ibc & 0 & 0 & 0 & 0 & 2 & 0 & 1 & 3 & 3 & 9 \\
  II & 0 & 1 & 1 & 1 & 11 & 6 & 12 & 5 & 3 & 40 \\
  \\
  All & 6 & 4 & 3 & 6 & 22 & 11 & 29 & 12 & 9 & 102 \\
  \hline
  \end{tabular}
  \parbox{\hsize}{Among SNe II, 4 are of Type IIb (2 in Sb and 2 in Scd galaxies).
                  Due to the uncertainties in the progenitor nature of Type IIn SNe,
                  and often their misclassification \citep[e.g.][]{2012MNRAS.424.1372A,2014MNRAS.441.2230H},
                  we remove them from the sample.}
  \end{minipage}
\end{table}

In S0--Sd galaxies, all CC SNe and the vast majority of Type Ia SNe belong to the disc,
rather than the bulge component \citep[][hereafter Paper~III]{2016MNRAS.456.2848H}.
Therefore, for the selected 106 SNe in this restricted sample of edge-on galaxies,
we perform a visual inspection of the SNe positions on the SDSS images
to identify the SNe from the bulge population of host galaxies.
The result is that three Type Ia (1990G, 1993aj, and 2003ge)
and one Type Ib/c (2005E) SNe may belong to the bulge because of their location.
The three SNe Ia are clearly outside the host discs,
located far in the bulge population.
The Type Ib/c SN is also located far from the host galaxy disc
but it is a peculiar, calcium-rich SN whose nature is still
under debate and may have a different progenitor from typical
CC \citep[e.g.][]{2010Natur.465..322P}.
All these four SNe are excluded from the sample.

\begin{figure}
\begin{center}$
\begin{array}{@{\hspace{0mm}}c@{\hspace{3mm}}c@{\hspace{0mm}}}
\includegraphics[width=0.48\hsize]{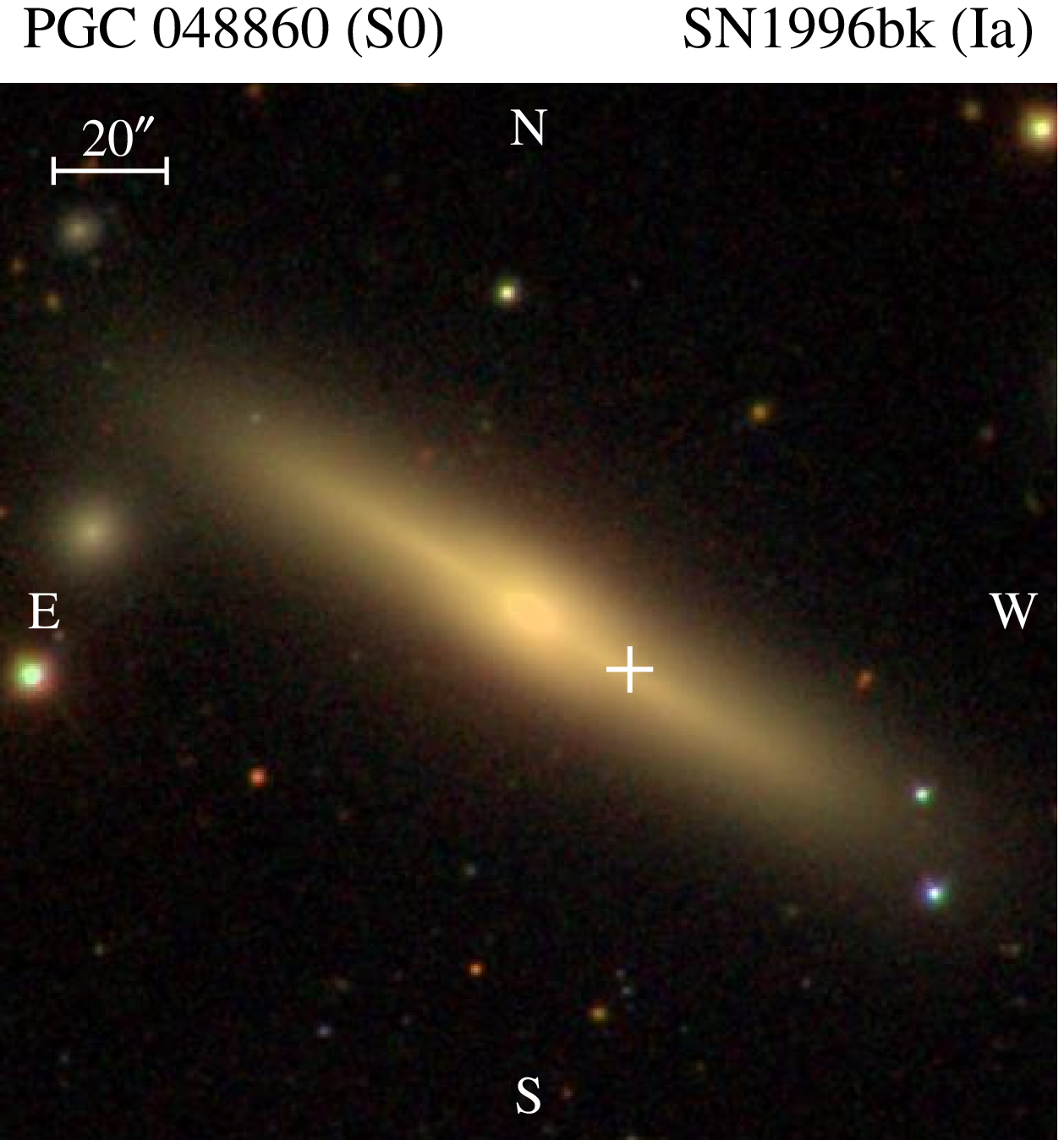} &
\includegraphics[width=0.48\hsize]{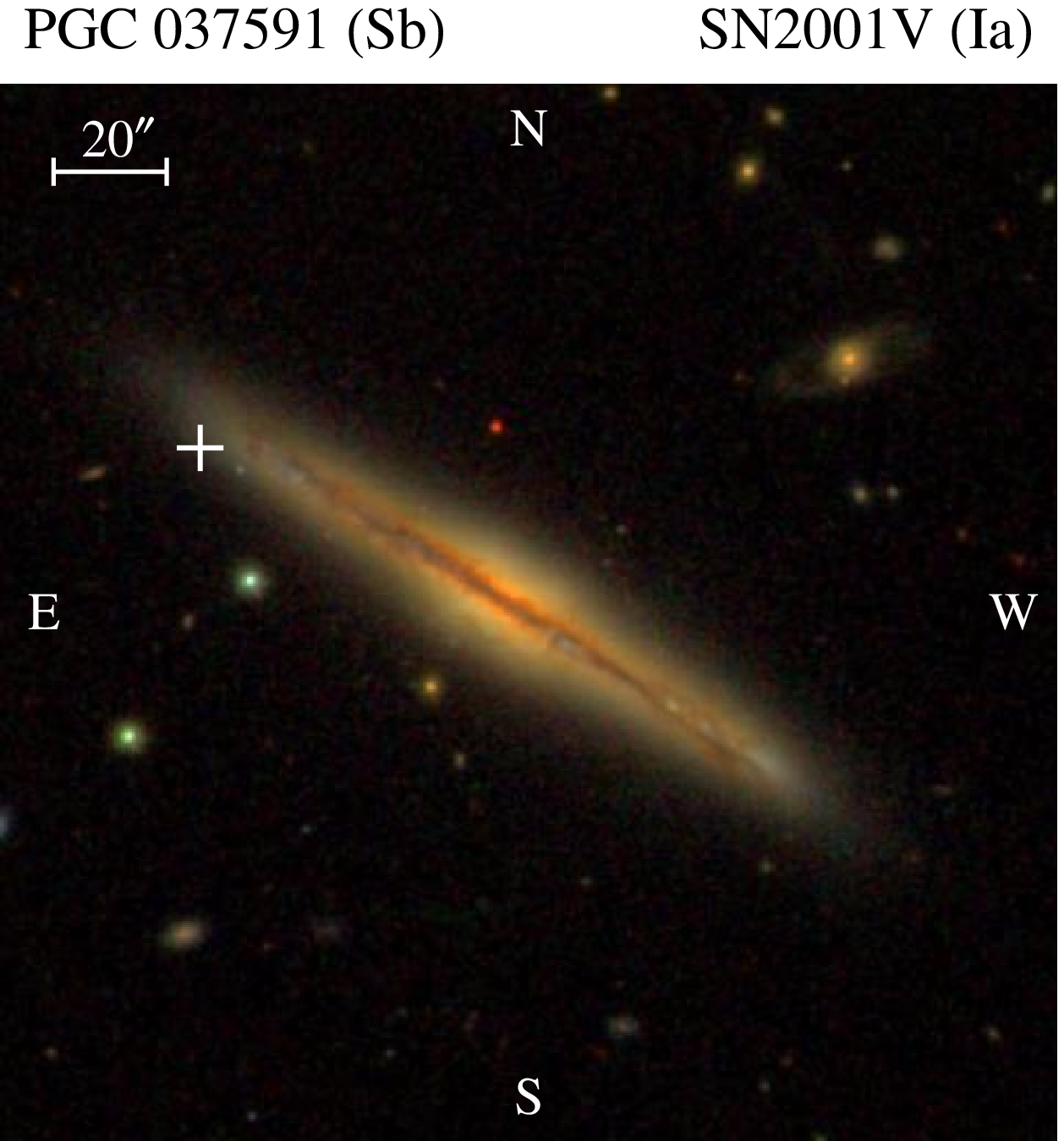}
\end{array}$
\end{center}
\begin{center}$
\begin{array}{@{\hspace{0mm}}c@{\hspace{3mm}}c@{\hspace{0mm}}}
\includegraphics[width=0.48\hsize]{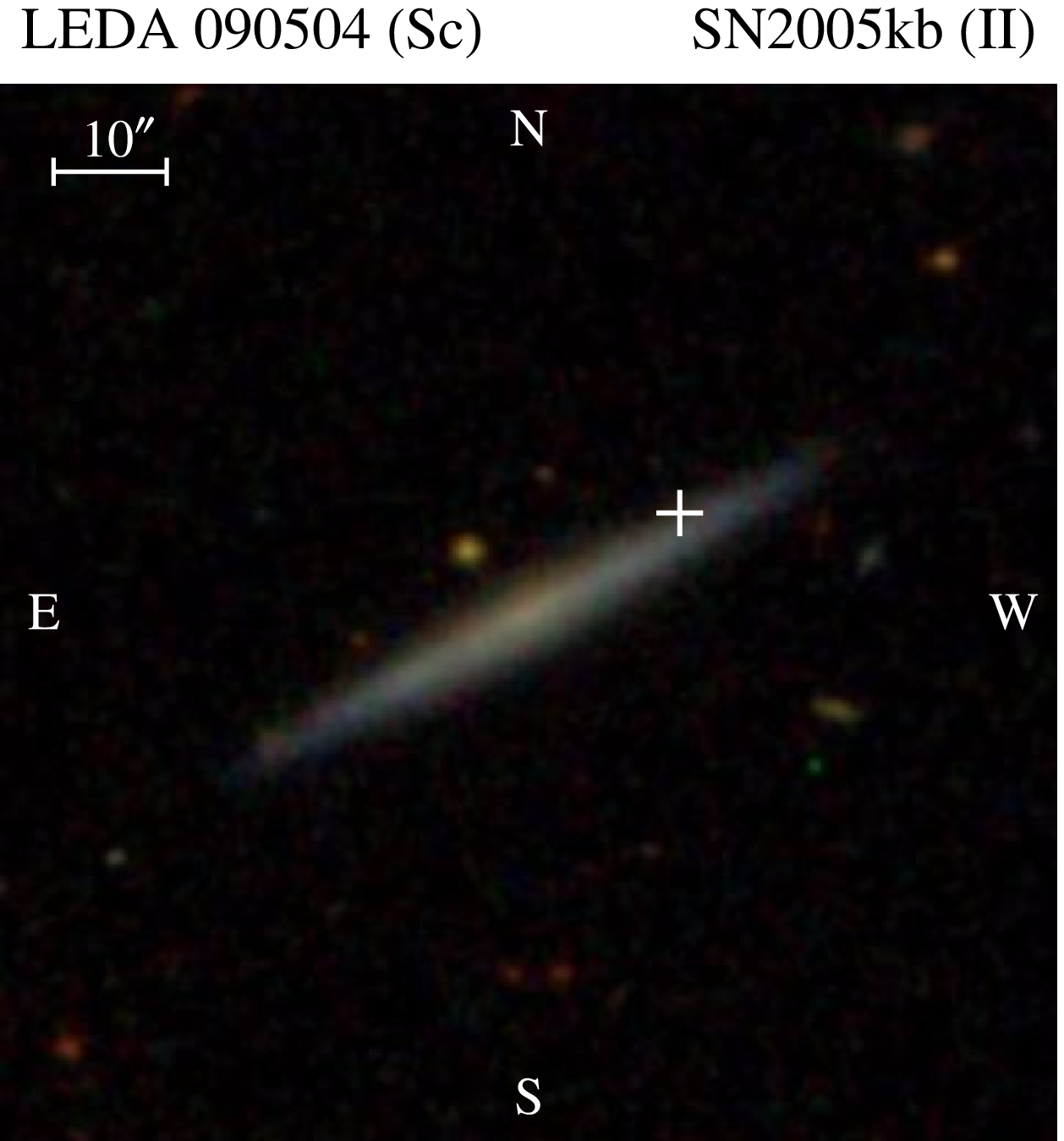} &
\includegraphics[width=0.48\hsize]{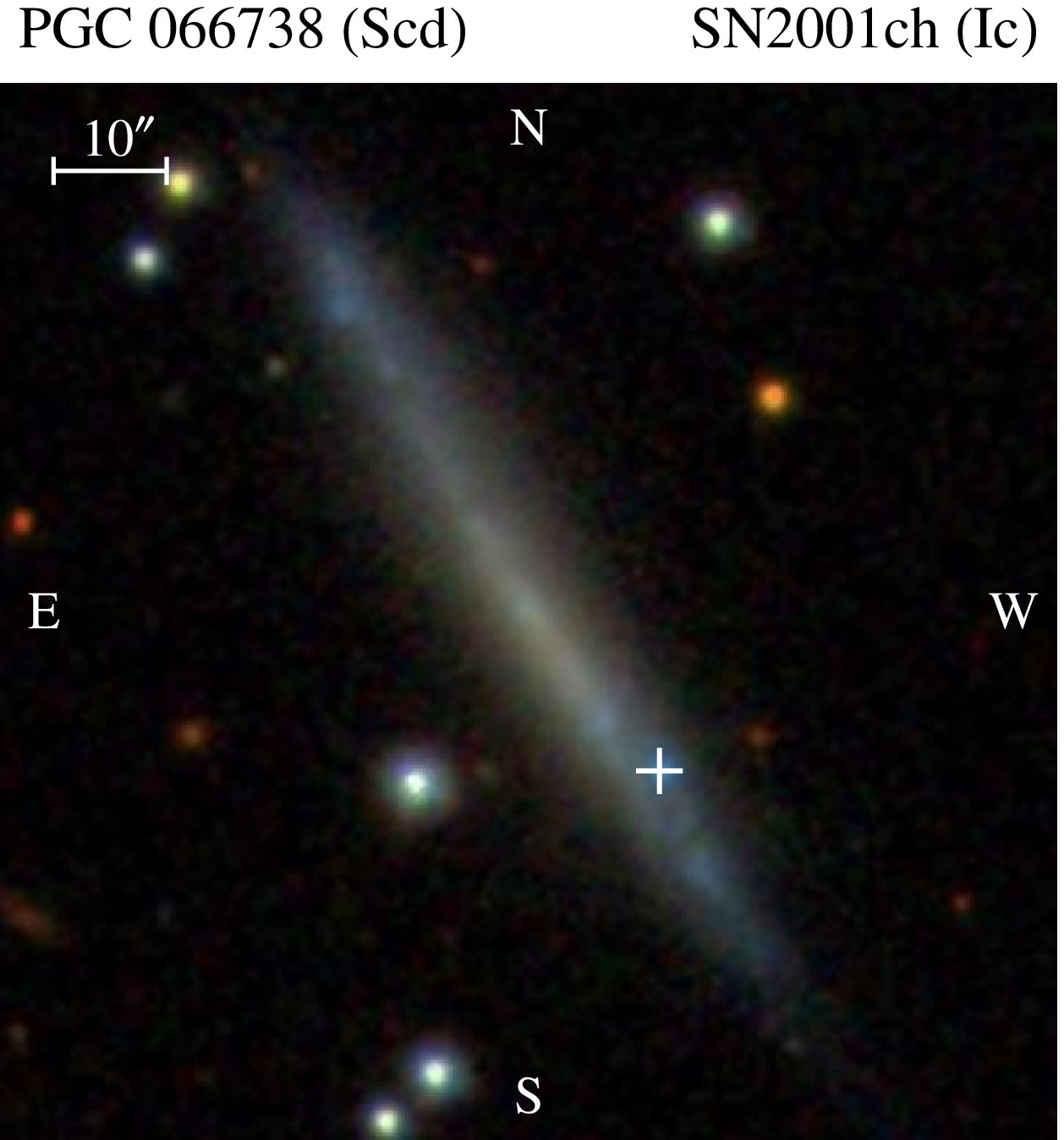}
\end{array}$
\end{center}
\caption{SDSS images representing examples of edge-on SNe host galaxies.
         The objects' identifiers with host morphologies and SN types
         (in parentheses) are listed at the top.
         The positions of SNe (marked by cross sign) are also shown.
         In all images, north is up and east to the left.}
\label{edgeon_exampl}
\end{figure}

After these restrictions, we are left with a sample of 102 SNe within 100 host galaxies.
The mean distance of this sample is $100\pm8$~Mpc,
the median distance and standard deviation are 78~Mpc and 84~Mpc, respectively.
The mean $D_{25}$ of our host galaxies is $108\pm10$ arcsec with the smallest value of 22 arcsec.
Table~\ref{table_SN_morph} displays the distribution of all SNe types
among the various considered morphological types of host galaxies.
Fig.~\ref{edgeon_exampl} shows images of typical examples of edge-on
host galaxies with marked positions of SNe.

\subsection{Measurements of the heights of SNe}
\label{vertH}

The heights of SNe above host galactic plane might be calculated
by using the simple formulas presented in \citet{2009A&A...508.1259H}
with available SNe offsets from host galaxy nuclei and PA of the galaxies
(see also Paper~\citetalias{2016MNRAS.456.2848H}).
However, as demonstrated in Paper~\citetalias{2012A&A...544A..81H},
SN catalogs report different offsets with different levels of accuracy.
Individual offsets are based on the determination of the positions of
the host galaxy nuclei, which might be uncertain and depend on many factors
(e.g. colour of image, plate saturation, galaxy peculiarity,
incorrect SDSS fiber targeting of the galaxy nucleus, etc.).
For more details, the reader is referred to Paper~\citetalias{2012A&A...544A..81H}.

\begin{figure}
\centering
\includegraphics[width=\hsize]{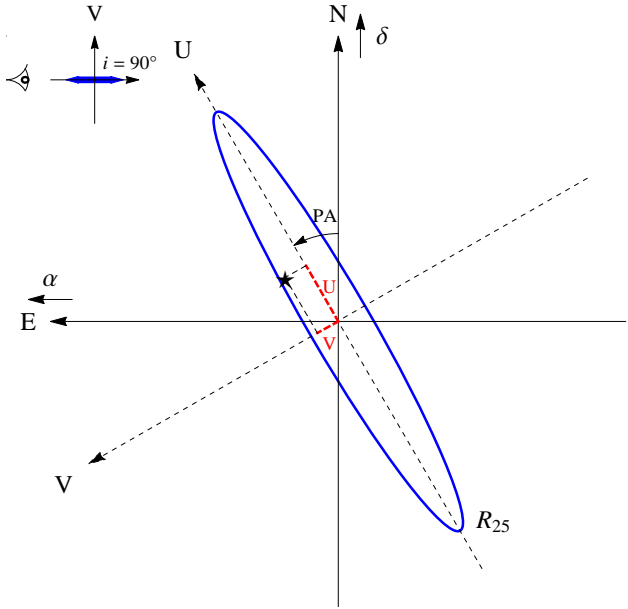}
\caption{Location of the SN within its edge-on host galaxy. The
         center of the galaxy is at the origin of coordinate systems
         and the asterisk is the projected location of the SN.
         The \textsc{u} (the projected galactocentric radius)
         and \textsc{v} (the height) are coordinates
         of the SN in host galaxy coordinate system
         along the major (U) and the minor (V) axes, respectively.
         The inset in the upper-left corner illustrates the $90^\circ$ inclination of
         the polar axis of the galaxy with respect to the line of sight.
\label{snloc}}
\end{figure}

For this study, using the SN coordinates and its edge-on host galaxy image
in the SDSS $g$-band, we measure the perpendicular distance, i.e., the height,
from the major axis of the fitted elliptical aperture of each galaxy to
the position of SN.
At the same time, using the coordinates of the host galaxy nucleus,
we also measure the projected galactocentric radius of SN along
the same major axis.\footnote{{\footnotesize We remind that
in comparison with the measured heights,
the measurements of projected galactocentric radii of SNe include
some minor inaccuracy because of the mentioned uncertain determination
of the exact point like positions of host galaxy nuclei.
The projected galactocentric radii are only used in
Fig.~\ref{VR25vsUR25} of Section~\ref{z0andhz} for ancillary purposes.}}
Fig.~\ref{snloc} schematically illustrates the geometrical location of an SN within an edge-on disc,
where \textsc{v} is the height (in arcsec) and
\textsc{u} is the projected galactocentric radius (in arcsec) of the SN.
A similar technique was also used in \citet{2016AstL...42..495P}
on the Digital Sky Survey (DSS) images
to determine the \textsc{v} and \textsc{u} coordinates of SNe.

\begin{figure}
\centering
\includegraphics[width=\hsize]{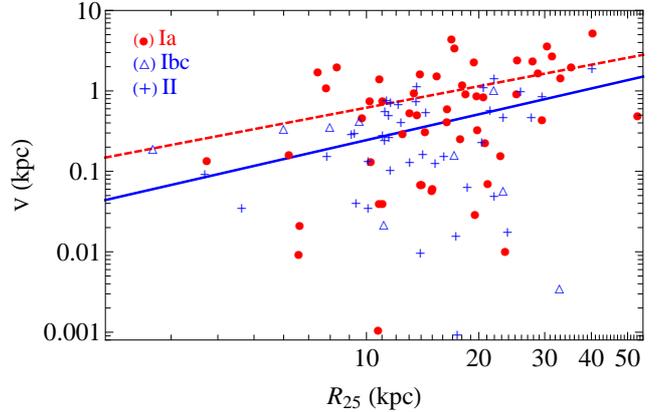}
\caption{Comparison of the heights \textsc{v} of SNe and $R_{25}$ of host galaxies in kpc.
         Red circles, blue triangles and crosses respectively show Types Ia, Ibc and II SNe.
         Red dashed (Ia) and blue solid (Ibc+II) lines are best fits to the samples.
\label{VverR25}}
\end{figure}

It is important to note that
as in the case of the radial distribution of SNe in face-on galaxies
\citep{2009A&A...508.1259H},
the distribution of linear distances in the vertical direction
is biased by the greatly different intrinsic sizes of
host discs. Fig.~\ref{VverR25} illustrates the comparison of
the heights \textsc{v} of SNe and $R_{25}$ of host galaxies in kpc.
Also shown are the best fit lines
\[
 \log(\textsc{v}_{\rm Ia}) = (-1.10\pm0.11) + (0.89\pm0.08) \, \log(R_{25}) \ ,
\]
\[
 \log(\textsc{v}_{\rm CC}) = (-1.68\pm0.15) + (1.07\pm0.13) \, \log(R_{25})
\]
with near unity slopes.
To check the significance of the correlations, we use the Spearman's rank correlation test,
which indicates strong positive trend between the heights and $R_{25}$ for Type Ia SNe
($r_{\rm s}=0.382, P=0.005$), while not significant for CC SNe ($r_{\rm s}=0.166, P=0.255$).
Therefore, in the remainder of this study, we use only relative heights
and projected galactocentric radii of SNe,
i.e., normalized to $R_{25}=D_{25}/2$ of host galaxies in $g$-band.

The full data base of 102 individual SNe
(SN designation, type, equatorial coordinates, \textsc{v} and \textsc{u}) and
their 100 host galaxies (galaxy SDSS designation, distance, morphological type
and corrected $g$-band $D_{25}$)
is available in the online version (Supporting Information) of this article.

\section{The model of stellar disc}
\label{discmodel}

In our model, the volumetric density
$\rho^{\rm SN}(\tilde{r}, \tilde{z})$ of SNe in the host axisymmetric stellar discs is assumed to
vary as follows in the radial $\tilde{r}$ and vertical $\tilde{z}$ directions:
\begin{equation}
\rho^{\rm SN}(\tilde{r}, \tilde{z})=\rho_0^{\rm SN} \exp(-\tilde{r}/\tilde{h}_{\rm SN})f(\tilde{z}) \ ,\\
\label{model1}
\end{equation}
where $\tilde{r}=R_{\rm SN}/R_{25}$,
$\tilde{z}=z_{\rm SN}/R_{25}$
and $(R_{\rm SN}, z_{\rm SN}\equiv\textsc{v})$ are cylindrical coordinates,
$\rho_0^{\rm SN}$ is the central volumetric density,
$\tilde h_{\rm SN} = h_{\rm SN}/R_{25}$ is the radial scale length, and $f(\tilde{z})$ is a function
describing the vertical distribution of SNe.

In equation~(\ref{model1}), we adopt a generalized vertical distribution
\begin{equation}
f(\tilde{z})={\rm sech}^{2/n}(n\tilde{z}/\tilde{z}_{0}^{\rm SN}) \ ,\\
\label{model2}
\end{equation}
where $\tilde{z}_{0}^{\rm SN} = z_{0}^{\rm SN}/R_{25}$ is
the vertical scale height of SNe and $n$ is a parameter controlling
the shape of the profile near the plane of host galaxy.
Following the vertical surface brightness distribution of edge-on galaxies
\citep[e.g.][]{1997A&A...327..966D,2002A&A...389..795B},
we also assume that the scale height of SNe is independent of
projected galactocentric radius
\citep[see also][for late-type galaxies]{1997A&A...320L..21D},
i.e., there is no disc flaring.

Recent photometric fits to the surface brightness distribution of a large number
of edge-on galaxies in near-infrared \citep[][]{2010MNRAS.401..559M}
and SDSS $g$-, $r$-, and $i$-bands (\citealt{2014ApJ...787...24B},
see also \citealt{2006AJ....131..226Y} for other photometric bands)
suggest that a value of $n=1$ is an appropriate model of stellar discs.
When $n \rightarrow \infty$, equation~(\ref{model2})
reduces to $f(\tilde{z})\sim \exp(-|\tilde{z}|/\tilde{H}_{\rm SN})$,
where $\tilde{H}_{\rm SN}=\tilde{z}_{0}^{\rm SN}/2$ at large heights,
and is widely used to successfully fit the dust distribution in edge-on galaxies
\citep[e.g.][]{2007A&A...471..765B,2014ApJ...787...24B}.

In linear units, the exponential (${\rm exp}$) form of $f(\tilde{z})$ is used
to model the distribution of Galactic stars \citep[e.g.][]{2001ApJ...553..184C,2003AJ....125.1958L},
novae \citep[e.g.][]{1997MNRAS.290..113H},
SNe \citep[e.g.][]{1994JRASC..88..369D,1997MNRAS.290..360H},
SN remnants \citep[e.g.][]{1972A&A....18..169I},
pulsars \citep[e.g.][]{2016Ap.....59...57A},
and extragalactic SNe \citep[e.g.][]{1997PhDT........11M,1998ApJ...502..177H,2016AstL...42..495P},
while the ${\rm sech}^{2}$ form is used to fit the vertical distribution of
resolved stars \citep{2005AJ....130.1574S} and
CC SNe \citep{2012MsT..........1M} in highly inclined nearby galaxies.

Note that ${\rm sech}^{2}$ profile ($n=1$) is expected for an isothermal stellar
population \citep[][]{1942ApJ....95..329S},
while ${\rm exp}$ profile ($n \rightarrow \infty$) can be obtained by
a combination of isothermal stellar populations
with different ``temperatures'' (velocity dispersions).
While at large heights, ${\rm sech}^{2} (x) \rightarrow 4\, {\rm exp}(-2x)$,
at low heights, the ${\rm sech}^{2}$ profile is uniform,
while the ${\rm exp}$ profile is cuspy.

\section{Results and discussion}
\label{resdiscus}

\subsection{The vertical distribution and scale height of SNe}
\label{z0andhz}

\begin{figure}
\begin{center}$
\begin{array}{@{\hspace{0mm}}c@{\hspace{0mm}}}
\includegraphics[width=1\hsize]{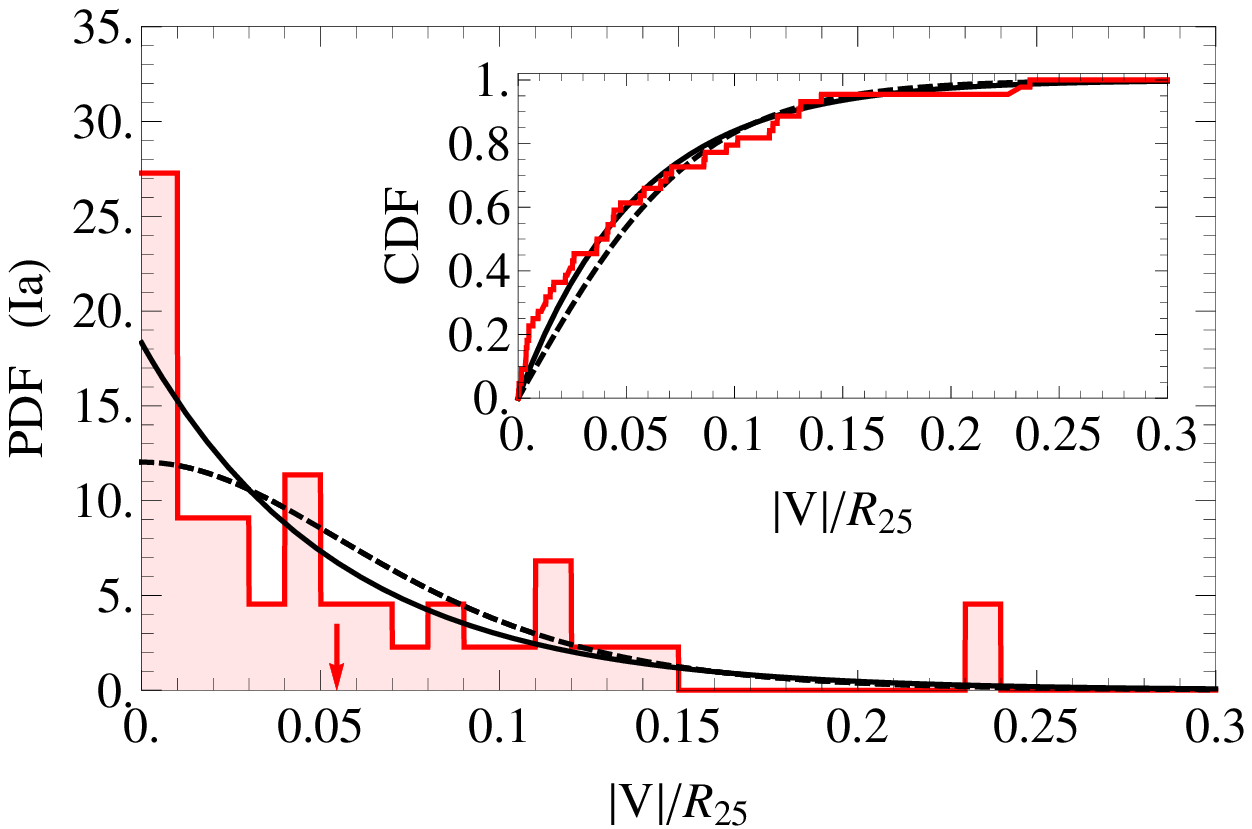}
\end{array}$
\end{center}
\begin{center}$
\begin{array}{@{\hspace{0mm}}c@{\hspace{0mm}}}
\includegraphics[width=1\hsize]{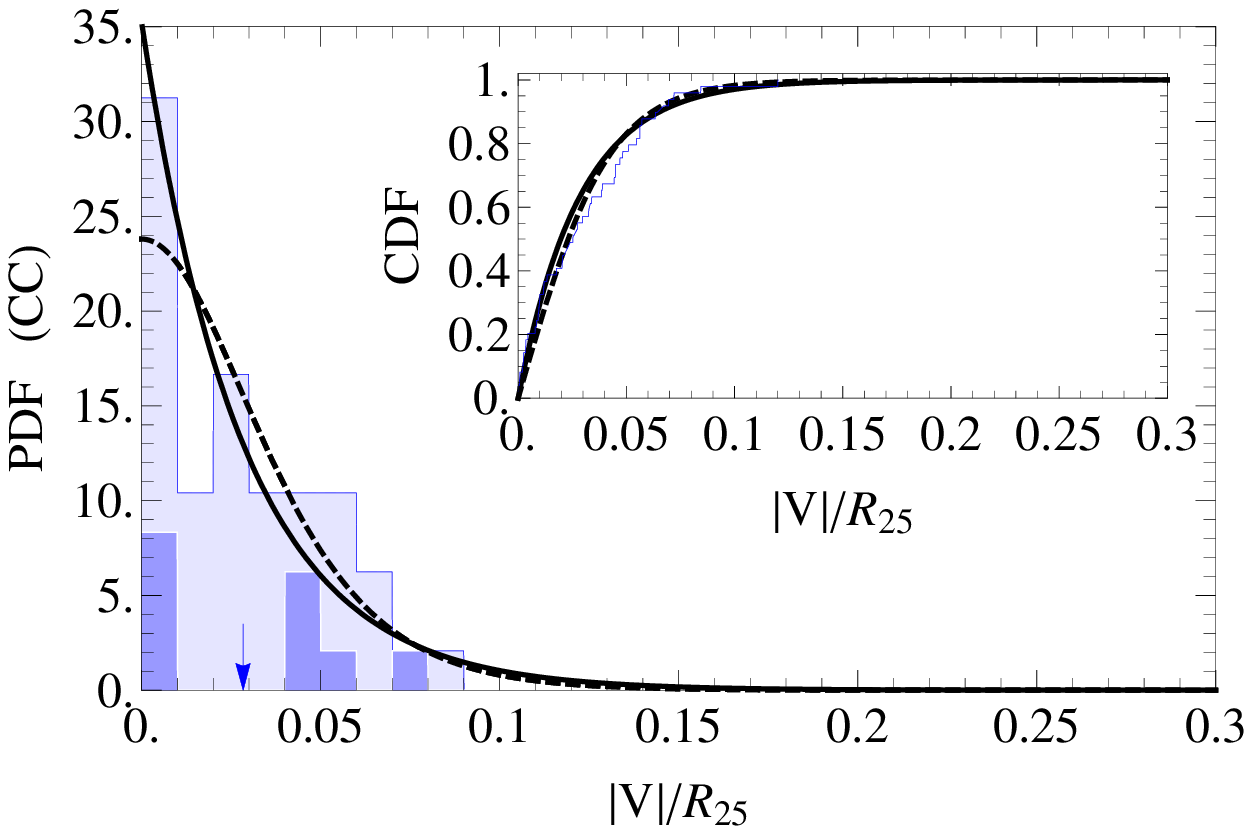}
\end{array}$
\end{center}
\caption{Vertical distribution of SNe (scaled to isophotal radius of disc) in Sa--Sd galaxies.
         Upper panel: fitted ${\rm sech}^{2}$ (dashed curve) and ${\rm exp}$ (solid curve) PDFs of
         the normalized absolute heights ($|\tilde{z}|\equiv|\textsc{v}|/R_{25}$)
         of Type Ia SNe (red histogram).
         Bottom panel: the same for CC SNe (blue histogram).
         The dark blue histogram presents the distribution of Type Ibc SNe only.
         The insets present the different forms of fitted CDFs
         in comparison with the SN distribution.
         The mean values of the distributions are shown by arrows.}
\label{VR25_distr}
\end{figure}

We fit ${\rm sech}^{2}$ and ${\rm exp}$ forms of $f(\tilde{z})$ profile to the
distribution of normalized absolute heights ($|\tilde{z}|\equiv|\textsc{v}|/R_{25}$) of SNe,
using maximum likelihood estimation (MLE).
Here, because of the small number statistics of Type Ibc SNe (see Table~\ref{table_SN_morph}),
we group them with Type II SNe in a larger CC SNe sample.
Fig.~\ref{VR25_distr} shows the histograms of the normalized heights
with the fitted ${\rm sech}^{2}$ and ${\rm exp}$ probability density functions (PDFs)
for Type Ia and CC SNe in Sa--Sd galaxies.\footnote{{\footnotesize For
this comparative illustration, we do not include S0--S0/a galaxies because
they host almost only Type Ia SNe (see Table~\ref{table_SN_morph}).
For the sake of visualization, the distribution of Type Ibc SNe is also presented in
the bottom panel of Fig.~\ref{VR25_distr}.}}
In columns~4, 7, and 10 of Table~\ref{tableallSNe},
we list the mean values of $|\tilde{z}|$ and
the maximum likelihood scale heights for both types of SNe in
various subsamples of host galaxies.

From column~4 of Table~\ref{tableallSNe}, it immediately becomes clear
that in all the subsamples of host galaxies the vertical distribution of CC SNe
is about twice closer to the plane of host disc than the distribution of Type Ia SNe.
In fact, the two-sample Kolmogorov--Smirnov (KS) and Anderson--Darling (AD)
tests,\footnote{{\footnotesize The two-sample AD test is more powerful than the KS test
\cite[][]{Engmann+11}, being more sensitive to differences in the tails of distributions.
Traditionally, we chose the threshold of 5 per cent for significance levels of the different tests.}}
shown in Table~\ref{diffSNe_KS_AD},
indicate that this difference is statistically significant
in Sa--Sd galaxies, although not significant if only late-type hosts are considered.

\begin{figure*}
\centering
\includegraphics[width=0.95\hsize]{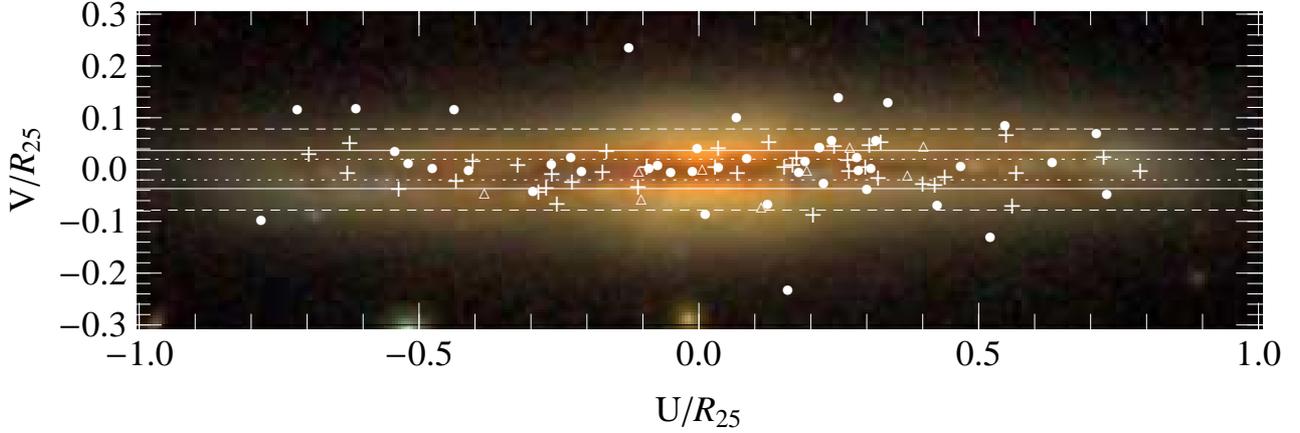}
\caption{Distribution of coordinates of SNe along the major ($\textsc{u}/R_{25}$)
         and minor axes ($\tilde{z}\equiv\textsc{v}/R_{25}$) of their Sa--Sd host galaxies.
         Circles, triangles and crosses respectively show Types Ia, Ibc and II SNe.
         One-sigma intervals of the distributions of the $\tilde{z}$ coordinates
         for Type Ia and CC (Ibc+II) SNe are presented by
         dashed ($\sigma=0.078$) and solid ($\sigma=0.037$) lines, respectively.
         Background SDSS image shows the PGC~037591 galaxy (scaled to the distribution),
         which is one of the representatives of the edge-on galaxies
         with a prominent dust line along the major axis.
         Dotted lines show the $|\tilde{z}|\leq0.02$ opaque region.
\label{VR25vsUR25}}
\end{figure*}

Note that four Type IIb SNe are included in Type II SNe sample (see Table~\ref{table_SN_morph}).
For Sa--Sd galasies, it might be reasonable also to group Types IIb and Ibc SNe
as a wider `stripped-envelop' (SE) SN class (13 objects) and
compare them with pure Type II SNe (35 objects).
However, we find no difference between the vertical distributions of
SE and pure Type II SNe ($P_{\rm KS}=0.401$, $P_{\rm AD}=0.320$),
resulting in statistically indistinguishable scale lengths between these SN types.
Therefore, in the remainder of this study, we will group all these subtypes as
the main CC SN sample and compare that with Type Ia SN sample.

\begin{table*}
  \centering
  \begin{minipage}{139mm}
  \caption{Consistency and scale heights of the vertical distributions of Type Ia and CC SNe
           in edge-on galaxies with ${\rm sech}^{2}$ ($n=1$) and ${\rm exp}$ ($n \rightarrow \infty$) models.}
  \tabcolsep 5.5pt
  \label{tableallSNe}
  \begin{tabular}{llrc@{\hspace{5mm}}ccc@{\hspace{5mm}}ccc}
  \hline
    \multicolumn{1}{c}{}&&&&\multicolumn{3}{c}{$n=1$}&\multicolumn{3}{c}{$n \rightarrow \infty$}\\
    \multicolumn{1}{l}{Host}&\multicolumn{1}{c}{SN}&\multicolumn{1}{c}{$N_{\rm SN}$}&\multicolumn{1}{c}{$\langle|\tilde{z}|\rangle$}&\multicolumn{1}{c}{$P_{\rm KS}$}&\multicolumn{1}{c}{$P_{\rm AD}$}&\multicolumn{1}{c}{$\tilde{z}_{0}^{\rm SN}$}&\multicolumn{1}{c}{$P_{\rm KS}$}&\multicolumn{1}{c}{$P_{\rm AD}$}&\multicolumn{1}{c}{$\tilde{H}_{\rm SN}$}\\
    \multicolumn{1}{l}{(1)}&\multicolumn{1}{c}{(2)}&\multicolumn{1}{c}{(3)}&
    \multicolumn{1}{c}{(4)}&\multicolumn{1}{c}{(5)}&\multicolumn{1}{c}{(6)}&\multicolumn{1}{c}{(7)}&
    \multicolumn{1}{c}{(8)}&\multicolumn{1}{c}{(9)}&\multicolumn{1}{c}{(10)}\\
  \hline
    S0--Sd&Ia&53&$0.058\pm0.009$&0.068&\textbf{0.012}&$0.089\pm0.015$&0.196&0.165&$0.058\pm0.009$\\
    \\
    Sa--Sd&Ia&44&$0.055\pm0.009$&0.147&\textbf{0.031}&$0.083\pm0.012$&0.319&0.239&$0.055\pm0.007$\\
    Sa--Sd&CC&48&$0.028\pm0.003$&0.644&0.209&$0.042\pm0.004$&0.648&0.287&$0.028\pm0.003$\\
    Sa--Sd$^\dag$&Ia&28&$0.082\pm0.011$&0.983&0.973&$0.098\pm0.014$&0.970&0.973&$0.062\pm0.012$\\
    Sa--Sd$^\dag$&CC&28&$0.044\pm0.003$&0.459&0.723&$0.041\pm0.006$&0.331&0.525&$0.024\pm0.004$\\
    \\
    Sa--Sbc&Ia&21&$0.061\pm0.014$&0.168&0.055&$0.094\pm0.014$&0.371&0.151&$0.061\pm0.011$\\
    Sa--Sbc&CC&21&$0.028\pm0.004$&0.860&0.299&$0.040\pm0.005$&0.492&0.239&$0.028\pm0.003$\\
    Sc--Sd&Ia&23&$0.049\pm0.011$&0.627&0.354&$0.073\pm0.018$&0.849&0.919&$0.049\pm0.009$\\
    Sc--Sd&CC&27&$0.029\pm0.005$&0.493&0.353&$0.044\pm0.007$&0.497&0.684&$0.029\pm0.004$\\
    \\
    Sb--Sc&Ia&30&$0.064\pm0.011$&0.476&0.212&$0.096\pm0.016$&0.679&0.482&$0.065\pm0.012$\\
    Sb--Sc&CC&32&$0.028\pm0.004$&0.476&0.203&$0.042\pm0.007$&0.586&0.264&$0.028\pm0.003$\\
    Sb--Sc$^\dag$&Ia&21&$0.089\pm0.013$&0.853&0.962&$0.108\pm0.021$&0.594&0.821&$0.070\pm0.014$\\
    Sb--Sc$^\dag$&CC&19&$0.044\pm0.004$&0.908&0.948&$0.041\pm0.008$&0.728&0.794&$0.024\pm0.006$\\
    Sb--Sc$^*$&Ia&24&$0.065\pm0.014$&0.686&0.281&$0.097\pm0.020$&0.927&0.657&$0.065\pm0.014$\\
    Sb--Sc$^*$&CC&31&$0.028\pm0.004$&0.422&0.224&$0.042\pm0.007$&0.576&0.335&$0.028\pm0.004$\\
  \hline
  \end{tabular}
  \parbox{\hsize}{\emph{Notes.}
    Columns~1 and 2 give the subsample;
    Col.~3 is the number of SNe in the subsample;
    Col.~4 is the mean of normalized absolute vertical distribution with
    the error of the mean;
    Cols.~5 and 6 are the $P_{\rm KS}$ and $P_{\rm AD}$ probabilities from one-sample KS and AD tests,
    respectively, that the vertical distribution of SNe is drawn from the best-fitting ${\rm sech}^{2}$ profile;
    Col.~7 is the maximum likelihood value of the scale height
    with bootstrapped error (repeated $10^3$ times);
    Cols.~8, 9, and 10 are, respectively, the same as Cols.~5, 6, and 7, but for
    the best-fitting ${\rm exp}$ profile.
    The subsamples labeled with `$\dag$' symbols correspond to SNe with $|\tilde{z}|>0.02$.
    The subsamples labeled with `$*$' symbols correspond to SNe with distances $\leq 200$~Mpc.
    We calculate the $P_{\rm KS}$ and $P_{\rm AD}$ using the calibrations by
    \citet{Massey51} and \citet{1986gft..book.....D}, respectively.
    The statistically significant deviations from the best-fitting profile are highlighted in bold.}
  \end{minipage}
\end{table*}
\begin{table}
  \centering
  \begin{minipage}{84mm}
  \caption{Comparison of the normalized absolute vertical distributions
           ($|\tilde{z}|\equiv|\textsc{v}|/R_{25}$) of SNe among different pairs of subsamples.}
  \tabcolsep 3pt
  \label{diffSNe_KS_AD}
  \begin{tabular}{llr@{\hspace{7mm}}c@{\hspace{7mm}}llr@{\hspace{7mm}}cc}
  \hline
    \multicolumn{3}{c}{Subsample~1}&&\multicolumn{3}{c}{Subsample~2}&&\\
    \multicolumn{1}{l}{Host}&\multicolumn{1}{c}{SN}&\multicolumn{1}{c}{$N_{\rm SN}$}&&\multicolumn{1}{l}{Host}&\multicolumn{1}{c}{SN}&\multicolumn{1}{c}{$N_{\rm SN}$}&\multicolumn{1}{c}{$P_{\rm KS}$}&\multicolumn{1}{c}{$P_{\rm AD}$}\\
  \hline
    Sa--Sd&Ia&44&versus&Sa--Sd&CC&48&\textbf{0.045}&\textbf{0.025}\\
    Sa--Sd$^\dag$&Ia&28&versus&Sa--Sd$^\dag$&CC&28&\textbf{0.011}&\textbf{0.003}\\
    Sa--Sbc&Ia&21&versus&Sa--Sbc&CC&21&\textbf{0.041}&\textbf{0.037}\\
    Sc--Sd&Ia&23&versus&Sc--Sd&CC&27&0.690&0.310\\
  \\
    Sa--Sbc&Ia&21&versus&Sc--Sd&Ia&23&0.387&0.440\\
    Sa--Sbc&CC&21&versus&Sc--Sd&CC&27&0.765&0.802\\
  \\
    Sb--Sc&Ia&30&versus&Sb--Sc&CC&32&\textbf{0.039}&\textbf{0.009}\\
    Sb--Sc$^\dag$&Ia&21&versus&Sb--Sc$^\dag$&CC&19&\textbf{0.013}&\textbf{0.001}\\
    Sb--Sc$^*$&Ia&24&versus&Sb--Sc$^*$&CC&31&0.112&\textbf{0.028}\\
  \hline
  \end{tabular}
  \parbox{\hsize}{\emph{Notes.} The subsamples labeled with `$\dag$' symbols correspond
                  to SNe with $|\tilde{z}|>0.02$.
                  The subsamples labeled with `$*$' symbols correspond to
                  SNe with distances $\leq 200$~Mpc.
                  The $P_{\rm KS}$ and $P_{\rm AD}$ are the probabilities from two-sample KS and AD tests,
                  respectively, that the two distributions being compared are drawn from the same
                  parent distribution.
                  The $P_{\rm KS}$ and $P_{\rm AD}$ are calculated using the calibrations by
                  \citet{Massey51} and \citet{Pettitt76}, respectively.
                  The statistically significant differences between the
                  distributions are highlighted in bold.}
  \end{minipage}
\end{table}

It is important to note that dust extinction in edge-on SN host galaxies
might have an impact on our estimated scale heights.\footnote{{\footnotesize Another factor,
such as a deviation from perfectly edge-on orientation of the host discs,
may also affect our estimation of the scale heights, increasing them.
However, we are quite confident that our galaxies can vary by a few degrees only
from perfectly edge-on orientation (see Section~\ref{inclin}).
In addition, other authors have demonstrated that slight deviations from $i=90^\circ$
have minimal impact on the derived structural parameters of the vertical distributions
of different stellar populations \citep[e.g.][]{1997A&A...327..966D}.}}
In Paper~\citetalias{2012A&A...544A..81H}, we demonstrated that in general
there is a lack of SNe host galaxies with high inclinations,
which can be explained by a bias in the discovery of SNe
due to strong dust extinction \citep[e.g.][]{1997ASIC..486...77C},
particularly in edge-on hosts \citep[e.g.][]{2015MNRAS.446.3768H}.

The vertical distribution of dust in disc galaxies
has an exponential profile with about three times smaller scale height
in comparison with distribution of all stars
\citep[$H_{\rm stars}/H_{\rm dust}\approx3$,
e.g.][]{2007A&A...471..765B}.\footnote{{\footnotesize This value can vary from two to four,
depending, respectively, on early- and late-type
morphology of edge-on spiral galaxies \citep[e.g.][]{2014MNRAS.441..869D}.}}
Analysing the vertical distribution of the resolved stellar populations in nearby edge-on galaxies,
\citet{2005AJ....130.1574S} found that the dust
has negligible impact on the distribution parameters of stars
at $|z| \gtrsim H_{\rm dust}$ heights
\citep[for the edge-on surface brightness profiles of unresolved populations,
see e.g.][]{2007A&A...471..765B}.
Therefore, in Table~\ref{tableallSNe},
to check the impact of the dust extinction on the obtained scale heights,
we also estimate the distribution parameters considering the SNe in Sa--Sd galaxies
only at $|\tilde{z}|>\tilde{H}_{\rm dust}$ heights.
For the average dust scale height, we use $\tilde{H}_{\rm dust}=0.02$,
roughly considering that $H_{\rm dust} \approx H_{\rm Ia}/3$
\citep[see also][]{1992AJ....104..696D}.
In Fig.~\ref{VR25vsUR25}, we show the distribution of coordinates of SNe along
the major ($\textsc{u}/R_{25}$) and minor axes ($\tilde{z}\equiv\textsc{v}/R_{25}$)
of their Sa--Sd host galaxies with the
$|\tilde{z}|\leq0.02$ opaque region,
and for the sake of visualization, we scale the distribution to
the PGC~037591 galaxy (also shown in Fig.~\ref{edgeon_exampl}, better known as NGC~3987),
which is one of the representatives of the edge-on galaxies
with a prominent dust line along the major axis.

From columns~7 and 10 of Table~\ref{tableallSNe}
(the subsamples of Sa--Sd hosts labeled with `$\dag$' symbols),
despite the small number statistics (column~3), we see that
the extinction by dust near to the plane of host galaxies does not strongly bias
the estimated scale heights of SNe.
The scale height of CC SNe with $|\tilde{z}|>0.02$ is almost equal to that
with the $|\tilde{z}|\geq0$, while the scale height of Type Ia SNe with $|\tilde{z}|>0.02$
is only $\sim15$ per cent greater (still statistically insignificant)
than that with the $|\tilde{z}|\geq0$.
In the remainder of this study, we will generally use the scale heights of SNe without
height-truncation due to the small number statistics and insignificance of the effect,
however, if needed, we will emphasize the impact of the dust extinction on the scale heights.

To check whether the distribution of SN heights follows the best-fitting
profiles, we perform one-sample KS and AD tests
on the cumulative distribution of the normalized absolute heights ($|\tilde{z}|$),
where the ${\rm sech}^{2}$ and ${\rm exp}$ models have
$E(|\tilde{z}|) = {\rm tanh}(|\tilde{z}|/\tilde{z}_{0}^{\rm SN})$
and $E(|\tilde{z}|) = 1 - \exp(-|\tilde{z}|/\tilde{h}_{\rm z}^{\rm SN})$
cumulative distribution functions (CDFs), respectively.
Columns~5, 6, 8 and 9 of Table~\ref{tableallSNe} show the KS and AD probabilities
that the vertical distributions are drawn from the best fitting profile.
Cumulative distributions of the heights and CDFs of the fitted forms
for Type Ia and CC SNe in Sa--Sd galaxies
are presented in the insets of Fig.~\ref{VR25_distr}.

From columns~5, 6, 8 and 9 of Table~\ref{tableallSNe},
we see that the vertical distribution is consistent with both profiles
in most subsamples of Type Ia SNe and in all subsamples of CC SNe.
For Type Ia SNe in Sa--Sd (also in S0--Sd) galaxies, the vertical distribution is
consistent with the ${\rm exp}$ profile, but not with the ${\rm sech}^{2}$ one
(as seen in the AD statistic but only very marginally in the KS statistic).
When we separate SNe Ia between early- and late-type host galaxies,
the inconsistency vanishes with only barely AD test significance in early-type spirals
(see the $P_{\rm AD}$ value in column~6 of Table~\ref{tableallSNe} for SNe Ia in Sa--Sbc galaxies).
The $\langle|\tilde{z}|\rangle$ value (scale heights too) for SNe Ia is $\sim25$ per cent
greater in Sa--Sbc galaxies than that in Sc--Sd hosts
(although the difference is not significant, see Table~\ref{diffSNe_KS_AD}),
while for CC SNe this parameter has a nearly constant value in the mentioned subsamples.
This effect can be attributed to the earlier and wider morphological distribution of SNe Ia host galaxies
(from S0/Sa to Sd, see Table~\ref{table_SN_morph} and also
Papers~\citetalias{2012A&A...544A..81H} and \citetalias{2014MNRAS.444.2428H})
in comparison with CC SNe hosts, and the systematically thinner vertical distribution of
the host stellar population from early- to late-type discs
\citep[e.g.][]{1998MNRAS.299..595D,2006AJ....131..226Y,2014ApJ...787...24B}.

In the first attempts to estimate the mean value of the vertical coordinates of SNe,
\citet{1981SvAL....7..254T,1987SvA....31...39T} used the distribution of SN colour excesses
without precise information on their spectroscopic types and host galaxy morphology
in a sample of non-edge-on spirals.
No difference was found in the vertical distributions of Type I and II SNe
with indication that both types belong to the young population I.
However, the inclinations of host galaxies and the uncertain separation\footnote{{\footnotesize Type Ibc SNe
were labelled as `I pec' types during observations before 1986
and included in the sample of Type I SNe.}}
of SN types might be the reason for the similarity between the vertical distributions
of the mentioned SN types.
Using a similar colour excess data of the best photometrically studied
Type Ia SNe in late-type galaxies,
\citet{1992AJ....104..696D} showed that these SNe have a considerably broader
vertical distribution than the dust discs of their hosts
and concluded that SNe Ia are older than the old disc population.

Direct measurements of the heights of SNe and estimation of the scales of
their vertical distributions were performed only in a small number of cases
\citep[][]{1997PhDT........11M,2012MsT..........1M,2016AstL...42..495P}.
\citet{1997PhDT........11M} examined the offsets between the major axes of
a sample of highly inclined ($i \geq 60^\circ$) galaxies and
the SNe they hosted in an attempt to measure the scale heights of Type Ia and II SNe.
Unfortunately, the sample of such objects was quite small (66 galaxies),
especially when restricted to galaxies at $i \geq 75^\circ$,
which resulted in statistically indistinguishable vertical distributions (in kpc)
between the mentioned types of SNe.
\citet{2012MsT..........1M} used data from the ASC to study
the vertical distribution (in kpc) of 64 CC SNe in highly inclined ($i \geq 80^\circ$)
Sa--Sd host galaxies. He showed that the distribution can be well fitted by
a ${\rm sech}^{2}$ profile.
However, these studies only used linear scales
to estimate the vertical distribution of SNe.
This is somewhat undesirable because the absolute distribution of SN heights (in kpc)
is biased by the greatly different intrinsic sizes of host discs
(as already shown in Fig.~\ref{VverR25}).

Most recently, \citet{2016AstL...42..495P} studied the absolute (in kpc) and
relative (normalized to radius of host galaxy) vertical distributions of SNe
using a sample of 26 Type Ia, 8 Ibc, and 44 II SNe
in spiral host galaxies with $i \geq 85^\circ$.
They found that the distributions can be fitted by ${\rm exp}$ profiles
with scale heights $\tilde{H}_{\rm Ia}=0.030\pm0.006$,
$\tilde{H}_{\rm Ibc}=0.024\pm0.006$,
and $\tilde{H}_{\rm II}=0.029\pm0.005$.
The scale heights for Type Ibc and II SNe
are in good agreement with our $\tilde{H}_{\rm CC}=0.028\pm0.003$ in Sa--Sd galaxies,
while their scale height for Type Ia SNe is much smaller than our
$\tilde{H}_{\rm Ia}=0.055\pm0.007$ in the same morphological bin.
However, the direct comparison of the scale heights
obtained by \citeauthor{2016AstL...42..495P} with ours is difficult
because they used the DSS images for reduction of SNe host galaxies
without mentioning the photometric band
(we assume that they used $B$-band), while we use the SDSS $g$-band to normalize
the heights to the $25^{\rm th}$ magnitude isophotal semimajor axes of host galaxies.
On the other hand, we are not able to check the consistency between
the morphological distributions of edge-on galaxies hosting Type Ia and CC SNe
in their and our samples because morphological types were not provided by
\citeauthor{2016AstL...42..495P}.

To exclude any dependence of scale height of host stellar population on the morphological type,
we analyse the vertical distribution of SNe in the most populated morphological bins,
i.e., in the narrower Sb--Sc subsample
(see Table~\ref{table_SN_morph}).\footnote{{\footnotesize On the other hand,
by selecting these bins we reduce the possible contribution
by SNe Ia from central bulges of host galaxies, although the bulge contribution is
only up to 9 per cent of the total SN Ia population in
Sa--Sd host galaxies (Barkhudaryan et al. in preparation).}}
In addition, the Sb--Sc subsample is more suitable for comparison of
the estimated vertical scale heights of SNe with those of
different stellar populations of thick and thin discs of the MW galaxy
(see Section~\ref{hrandhz}), and to exclude a small number of
very thin discs \citep[see e.g.][]{2017MNRAS.465.3784B},
which usually appear in late-type galaxies.

\begin{figure}
\begin{center}$
\begin{array}{@{\hspace{0mm}}c@{\hspace{0mm}}}
\includegraphics[width=1\hsize]{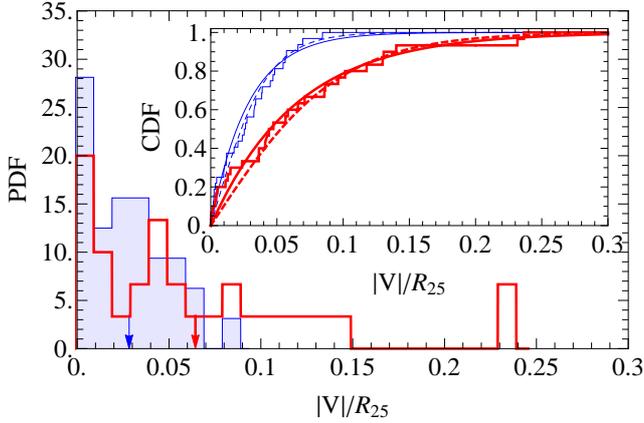}
\end{array}$
\end{center}
\caption{Vertical distributions of Type Ia (red thick line) and CC (blue thin line) SNe in Sb--Sc galaxies.
         The inset presents the cumulative distributions of SNe and fitted ${\rm sech}^{2}$ (dashed curve)
         and ${\rm exp}$ (solid curve) CDFs.
         The mean values of the distributions are shown by arrows.}
\label{VR25_distr_SbSc}
\end{figure}

From Table~\ref{tableallSNe}, we conclude that the vertical distributions of Type Ia and CC SNe
in Sb--Sc galaxies can be well fitted by both the ${\rm sech}^{2}$ and ${\rm exp}$ profiles.
The vertical distribution of CC SNe is significantly different from that
of Type Ia SNe (Table~\ref{diffSNe_KS_AD}), being $2.3\pm0.5$ times more concentrated
to the plane of the host disc (Table~\ref{tableallSNe}).
This difference also exists when the above-mentioned effect of the dust extinction
is considered for the particular subsample
(Sb--Sc hosts labeled with `$\dag$' symbols in Tables~\ref{tableallSNe} and \ref{diffSNe_KS_AD}).
In Fig.~\ref{VR25_distr_SbSc}, we present the comparison of vertical distributions
as well as the fitted ${\rm sech}^{2}$ and ${\rm exp}$ CDFs between
both the types of SNe in Sb--Sc host galaxies.

It is important to note that Type Ia SNe, because of their comparatively high luminosity
\citep[in about two absolute magnitudes in $B$-band, e.g.][]{2002AJ....123..745R}
and the presence of dedicated surveys,
are discovered at much greater distances than CC SNe (see Paper~\citetalias{2012A&A...544A..81H}).
To check the possible distance biasing on the vertical distribution of SNe,
we truncate the sample of Sb--Sc galaxies to distances $\leq 200$~Mpc.\footnote{{\footnotesize It would be
more effective to check this with distance-truncation at 150 (100) Mpc
(see Papers~\citetalias{2014MNRAS.444.2428H} and \citetalias{2016MNRAS.456.2848H}),
however the remaining statistics in this case is very low, which destroys any comparison with significance.
With the mentioned distance-truncation, we have only 19 (9) Type Ia SNe
with $\langle|\tilde{z}|\rangle=0.071\pm0.019$ ($0.086\pm0.025$) and 30 (24) CC SNe
with $\langle|\tilde{z}|\rangle=0.027\pm0.005$ ($0.031\pm0.006$).}}
In Table~\ref{tableallSNe}, the comparison of $\langle|\tilde{z}|\rangle$,
$\tilde{z}_{0}^{\rm SN}$, and $\tilde{H}_{\rm SN}$
as well as $P_{\rm KS}$ and $P_{\rm AD}$ values of distance-truncated sample (labeled with `$*$' symbols)
with those of Sb--Sc host galaxies allows to conclude that possible distance biasing
in our sample is negligible.
Due to the smaller number statistics, we get larger error bars in Table~\ref{tableallSNe},
and lose only the KS test significance in Table~\ref{diffSNe_KS_AD}.
Therefore, in the remainder of this study, we will use
SNe in Sb--Sc galaxies without distance-truncation.

\subsection{The thick and thin discs}
\label{hrandhz}

It is largely accepted that the disc of the MW,
one of the well-studied representatives of Sb--Sc classes,
is separated into at least three components/populations:
(1) the youngest star-forming disc ($\tilde{H}\lesssim0.01$),
including molecular clouds and massive young stars;
(2) the younger thin disc ($\tilde{H}\sim0.02$), which contains stars with a wide range of ages;
and (3) the old thick disc ($\tilde{H}\sim0.06$), composed almost exclusively of older stars
\citep[][]{1983MNRAS.202.1025G,1996AA...305..125R,1997AA...324...65N,1999AA...348...98B,2001MNRAS.322..426O,
2001ApJ...553..184C,2003AJ....125.1397C,2003AJ....125.1958L,2008ApJ...673..864J,2016AstL...42....1B}.
For extragalactic discs of nearby edge-on spirals, the thick and thin components are also resolved
\citep[e.g.][]{2005AJ....130.1574S,2006AJ....131..226Y}.
In this sense, we may be able to put constraints on the nature of the progenitors
of Type Ia and CC SNe by comparing the parameters of their distributions
($\tilde{H}_{\rm SN}$ or $\tilde{z}_{0}^{\rm SN}$ and
$h_{\rm SN}/z_{0}^{\rm SN}$ or $h_{\rm SN}/H_{\rm SN}$)
in edge-on Sb--Sc galaxies with those of different stellar populations
of thick and thin discs of MW and other similar galaxies.
Note that the mean luminosity of our sample of Sb--Sc host galaxies
($\langle M_{\rm g} \rangle=-20.5\pm1.0$)
is in good agreement with that of the MW
\citep[$\langle M_{\rm g}^{\rm MW} \rangle=-21.0\pm0.5$,][]{2015ApJ...809...96L}.

In Table~\ref{scaleheightsvsothers}, we list the ${\rm exp}$ scale heights of SNe estimated in this study
and the ${\rm exp}$ scale heights of the MW thick and thin discs derived from star counts
(from hundreds of thousands to millions of individual stars) by other authors.
As can be seen, the scale height of the vertical distribution of CC SNe is consistent with those of
younger stellar population in the thin disc
\citep[a wide range of ages  up to a few Gyr,][]{2011ApJ...737....8L},
while the scale height of Type Ia SNe is consistent with those of old population in
the thick disc \citep[from a few Gyr up to $\sim10$~Gyr,][]{2011ApJ...737....8L} of the MW galaxy.

\begin{table}
  \centering
  \begin{minipage}{77mm}
  \caption{Comparison of the $\tilde{H}_{\rm SN}$ values of Type Ia and CC SNe
           in edge-on Sb--Sc galaxies with those of the MW thick and thin discs.}
  \tabcolsep 4pt
  \label{scaleheightsvsothers}
  \begin{tabular}{lrl}
  \hline
    \multicolumn{1}{l}{Host}&\multicolumn{1}{c}{$\tilde{H}$}&\multicolumn{1}{l}{Reference}\\
  \hline
    MW thin disc&0.020 $\pm$ 0.005&\citet{2008ApJ...673..864J}\\
    MW thin disc&0.022 $\pm$ 0.003&\citet{2001ApJ...553..184C}\\
    MW thin disc&0.022 $\pm$ 0.005&\citet{2003AJ....125.1958L}\\
    \textbf{SNe CC (Sb--Sc)}&\textbf{0.028} $\pm$ \textbf{0.003}&\textbf{This study}\\
\\
    MW thick disc&0.050 $\pm$ 0.005&\citet{2001ApJ...553..184C}\\
    MW thick disc&0.051 $\pm$ 0.005&\citet{1996AA...305..125R}\\
    MW thick disc&0.057 $\pm$ 0.014&\citet{2001MNRAS.322..426O}\\
    MW thick disc&0.058 $\pm$ 0.005&\citet{2003AJ....125.1958L}\\
    MW thick disc&0.060 $\pm$ 0.013&\citet{2008ApJ...673..864J}\\
    MW thick disc&0.061 $\pm$ 0.020&\citet{1999AA...348...98B}\\
    \textbf{SNe Ia (Sb--Sc)}&\textbf{0.065} $\pm$ \textbf{0.012}&\textbf{This study}\\
    MW thick disc&0.067 $\pm$ 0.008&\citet{1997AA...324...65N}\\
  \hline
  \end{tabular}
  \parbox{\hsize}{\emph{Notes.} The MW $\tilde{H}$ values are calculated
                  using the original values of $H$ from the references and
                  assuming $R_{25}^{\rm MW}=15\pm1$~kpc.
                  The $\tilde{H}$ values are listed in ascending order.}
  \end{minipage}
\end{table}

Note that, in Table~\ref{scaleheightsvsothers}, the MW $\tilde{H}$ values
are calculated using the original values of $H$ (in kpc) from the references
and assuming $R_{25}^{\rm MW}=15\pm1$~kpc, i.e., $\tilde{H}=H/R_{25}^{\rm MW}$,
while the ratio of radial to vertical scales ($h/H$)
would be better for a comparison of SNe distribution with the distribution of stars in the MW,
avoiding the use of ambiguous value of $R_{25}^{\rm MW}$.

In Paper~\citetalias{2016MNRAS.456.2848H}, we studied the radial distributions of SNe and
estimated the scale lengths of Type Ia and CC SNe using a well-defined sample of
500 nearby SNe and their low-inclined ($i \leq 60^\circ$) and morphologically
non-disturbed S0--Sm host galaxies from the SDSS.\footnote{{\footnotesize At these
inclinations, dust extinction has minimal impact on the efficiency of
SNe discovery \citep[e.g.][]{1997ASIC..486...77C}, making the estimation of
the scale lengths as the most reliable.}}
In particular, the radial distributions of Type Ia and CC SNe in spiral galaxies
are consistent with one another and with an exponential surface density
according to $\exp(-\tilde{r}/\tilde{h}_{\rm SN})$ in equation~(\ref{model1})
where $\tilde{r}=R_{\rm SN}/R_{25}$ and
$\tilde{h}_{\rm SN}=h_{\rm SN}/R_{25}=0.21\pm0.02$.
However, to be consistent with the present study,
we use the estimation of the scale lengths of SNe
restricted to Sb--Sc host galaxies from that sample.
Note that the similar determination of the sample of the present paper is not possible
because of its extreme inclination.
For both types of SNe, we find $\tilde{h}_{\rm SN}=0.20\pm0.02$
using 79 Type Ia and 198 CC SNe.

\begin{table}
  \centering
  \begin{minipage}{76mm}
  \caption{Comparison of the length/height ratios of
           Type Ia and CC SNe in Sb--Sc galaxies with those of the MW stars in the thick and thin discs.}
  \tabcolsep 4pt
  \label{hhzvsothers}
  \begin{tabular}{lrl}
  \hline
    \multicolumn{1}{l}{Host}&\multicolumn{1}{c}{$h/H$}&\multicolumn{1}{l}{Reference}\\
  \hline
    \textbf{SNe Ia (Sb--Sc)}&\textbf{3.08} $\pm$ \textbf{0.65}&\textbf{This study}\\
    MW thick disc&3.30 $\pm$ 1.97&\citet{1999AA...348...98B}\\
    MW thick disc&3.68 $\pm$ 1.08&\citet{1996AA...305..125R}\\
    MW thick disc&4.00 $\pm$ 1.13&\citet{2008ApJ...673..864J}\\
    MW thick disc&4.30 $\pm$ 1.29&\citet{2001MNRAS.322..426O}\\
    MW thick disc&4.50 $\pm$ 0.46&\citet{1997AA...324...65N}\\
    MW thick disc&5.41 $\pm$ 0.41&\citet{2003AJ....125.1958L}\\
  \\
    MW thin disc&6.82 $\pm$ 3.03&\citet{2001ApJ...553..184C}\\
    \textbf{SNe CC (Sb--Sc)}&\textbf{7.14} $\pm$ \textbf{1.05}&\textbf{This study}\\
    MW thin disc&8.67 $\pm$ 2.45&\citet{2008ApJ...673..864J}\\
    MW thin disc&10.86 $\pm$ 2.70&\citet{2003AJ....125.1958L}\\
  \hline
  \end{tabular}
  \parbox{\hsize}{\emph{Notes.} For both the types of SNe,
                  we use $\tilde{h}_{\rm SN}=0.20\pm0.02$ (Paper~\citetalias{2016MNRAS.456.2848H}).
                  The $h/H$ values are listed in ascending order.}
  \end{minipage}
\end{table}

In Table~\ref{hhzvsothers}, we list the ratios of radial to vertical scales of SNe
($h_{\rm SN}/H_{\rm SN}$) estimated in this study
and the analogous ratios of MW thick and thin discs derived from star counts by other authors.
The ratio of scales of CC SNe appears consistent with those of the younger stellar population
in the thin disc, while the corresponding ratio of Type Ia SNe is consistent
with the old population in the thick disc of the MW (although on the small side).

It should be noted that the parameters of the vertical distributions of different
stellar populations in the MW
are determined using samples dominated by stars relatively near the Sun,
not including the sizable population of the disc
\citep[see the discussion in][]{2012ApJ...753..148B}.
Therefore, the structural parameters of the MW may
be different from those of other galaxies.
In particular, \citet{2005AJ....130.1574S} analysed the vertical distribution
of the resolved stellar populations in nearby six edge-on Sc galaxies observed with
the Hubble Space Telescope and found that the ratios of radial to vertical scales
of young star-forming discs are much smaller ($\sim$~3-4 times) than that of the MW.
In other words, the young star-forming discs of their sample galaxies
are much thicker in comparison with that of the MW.
Their results are in agreement with those of \citet{2006AJ....131..226Y},
who analysed the vertical structure of 34 late-type, edge-on, undisturbed disc galaxies
using the two-dimensional fitting to their photometric profiles.

Interestingly, \citet{2005AJ....130.1574S} found that the scale height of
a stellar population increases with age,
which is also correct for the MW galaxy \citep[e.g.][]{2003AJ....125.1397C,2012ApJ...753..148B}.
They used colour-magnitude diagrams (CMDs) to estimate the ages of resolved stellar populations
\citep[see figs.~1 and 4 in][]{2005AJ....130.1574S}.
The young population in their main-sequence (MS) box of the CMD is dominated by stars with
ages from $\sim10$~Myr up to $\sim100$~Myr,
the intermediate population in the asymptotic giant branch (AGB) box
is dominated by stars with ages from a few 100~Myr up to a few Gyr,
while the old population in the red giant branch (RGB) box is dominated by stars
with ages from a few Gyr up to $\sim10$~Gyr.
In light of this, we compare in Table~\ref{hz0vsothers} the ratios of radial to vertical scales
of SNe with those detected from resolved stars in nearby edge-on late-type galaxies
\citep[e.g.][]{2005AJ....130.1574S}
and from unresolved populations of extragalactic thick and thin discs
estimated using the edge-on surface brightness profiles
\citep[e.g.][]{2006AJ....131..226Y,2014ApJ...787...24B}.\footnote{{\footnotesize Here,
to be consistent with the original values from the references,
we use the $h_{\rm SN}/z_{0}^{\rm SN}$ ratios.}}

\begin{table*}
  \centering
  \begin{minipage}{111mm}
  \caption{Comparison of the length to ${\rm sech}^{2}$ height ratios of
           Type Ia and CC SNe in Sb--Sc galaxies with those detected from
           resolved stars in nearby edge-on galaxies
           and from unresolved populations of extragalactic thick and thin discs.}
  \tabcolsep 11pt
  \label{hz0vsothers}
  \begin{tabular}{lrl}
  \hline
    \multicolumn{1}{l}{Host}&\multicolumn{1}{c}{$h/z_{0}$}&\multicolumn{1}{l}{Reference}\\
  \hline
    Edge-on Sc galaxies$^a$ (RGB-box)&1.83 $\pm$ 0.99&\citet{2005AJ....130.1574S}\\
    \textbf{SNe Ia (Sb--Sc)}&\textbf{2.08} $\pm$ \textbf{0.40}&\textbf{This study}\\
    Edge-on Sc galaxies$^a$ (AGB-box)&2.40 $\pm$ 1.30&\citet{2005AJ....130.1574S}\\
    Edge-on galaxies$^b$ (thick$+$thin disc)&2.67 $\pm$ 0.86&\citet{2014ApJ...787...24B}\\
    Edge-on Sd galaxies$^c$ (thick disc)&2.87 $\pm$ 0.72&\citet{2006AJ....131..226Y}\\
  \\
    Edge-on Sc galaxies$^a$ (MS-box)&3.83 $\pm$ 1.79&\citet{2005AJ....130.1574S}\\
    \textbf{SNe CC (Sb--Sc)}&\textbf{4.76} $\pm$ \textbf{0.93}&\textbf{This study}\\
    Edge-on Sd galaxies$^c$ (thin disc)&5.48 $\pm$ 1.15&\citet{2006AJ....131..226Y}\\
  \hline
  \end{tabular}
  \parbox{\hsize}{\emph{Notes.} For both the types of SNe, we use $\tilde{h}_{\rm SN}=0.20\pm0.02$
                  (Paper~\citetalias{2016MNRAS.456.2848H}).
                  The $h/z_{0}$ values are listed in ascending order.\\
                  $^a$The mean ratio of all six galaxies with the additional components of
                  NGC~55 and NGC~4631 \citep[from table~4 in][]{2005AJ....130.1574S}.
                  These galaxies have lower masses than the MW.\\
                  $^b$To be consistent with the present study and the mentioned references,
                  the mean ratio in $g$-band is estimated for a subsample of 529 galaxies
                  from table~4 in \citet{2014ApJ...787...24B}
                  with bulge-to-total luminosity ratio (B/T) in $r$-band between 0.2 to 0.4 and
                  distances $\leq$ 200 Mpc (a few galaxies, with obviously incorrect B/T values, are removed).
                  The mean luminosity of this subsample ($\langle M_{\rm g} \rangle=-20.9\pm0.7$,
                  corrected for Galactic extinction) is in good agreement with that of our Sb--Sc host galaxies
                  ($\langle M_{\rm g} \rangle=-20.5\pm1.0$).\\
                  $^c$The mean ratio of all 34 galaxies in $R$-band from table~4 in \citet{2006AJ....131..226Y}.
                  These galaxies have lower kinematic masses than the MW.}
  \end{minipage}
\end{table*}

In Table~\ref{hz0vsothers}, we see that the ratio of scales of the distribution
of CC SNe is consistent with those of the resolved MS-box stars in \citet{2005AJ....130.1574S}
and unresolved stellar population of the thin disc in \citet{2006AJ....131..226Y}.
On the other hand, the $h_{\rm SN}/z_{0}^{\rm SN}$ ratio
of Type Ia SNe is consistent and located between
the values of the same ratios of resolved RGB- and AGB-box stars, respectively \citep{2005AJ....130.1574S}.
In addition, the $h_{\rm SN}/z_{0}^{\rm SN}$ ratio of Type Ia SNe
is consistent with those of the unresolved population of the thick disc in \citet{2006AJ....131..226Y}
and with the thick$+$thin disc population in \citet{2014ApJ...787...24B}.

These results are in good agreement with the age-scale height relation
of stars in galaxy discs \citep[e.g.][]{2005AJ....130.1574S,2012ApJ...753..148B},
and that Type Ia SNe result from stars of different ages
\citep[from $\sim0.5$~Gyr up to $\sim10$~Gyr, see][]{2012PASA...29..447M},
with even the shortest lifetime progenitors having much longer lifetime than
the progenitors of CC SNe \citep[from a few Myr up to $\sim0.2$~Gyr, see][]{2017A&A...601A..29Z}.

\section{Conclusions}
\label{concl}

In this fifth paper of a series, using a well-defined and homogeneous sample of
SNe and their edge-on host galaxies from the coverage of SDSS DR12,
we analyse the vertical distributions and estimate
the ${\rm sech}^{2}$ and ${\rm exp}$ scale heights
of the different types of SNe, associating them to
the thick or thin disc populations of galaxies.
Our sample consists of 100 nearby (the mean distance is $100\pm8$~Mpc),
high-inclination ($i \geq 85^\circ$), and morphologically non-disturbed S0--Sd galaxies,
hosting 102 SNe in total.

The extinction by dust near to the plane of edge-on host galaxies
has an insignificant impact on our estimated SN scale heights, although
as was shown previously (e.g. Paper~\citetalias{2012A&A...544A..81H}),
it is significantly decreasing the efficiency
of SN discovery in these galaxies.
We also check that there is no strong redshift bias within our SNe and host galaxies samples,
which could drive the observed behaviours of the vertical distributions of
the both SN types in host galaxies with edge-on discs.

The results obtained in this article are summarized below,
along with their interpretations.

\begin{enumerate}
\item For the first time, we show that in both early- and late-type edge-on spiral galaxies
      the vertical distribution of CC SNe is about twice more concentrated to the plane of host disc than
      the distribution of Type Ia SNe (Fig.~\ref{VR25_distr} and Table~\ref{tableallSNe}).
      The difference between the distributions of the SN types is statistically significant
      with only the exception in late-type hosts (Table~\ref{diffSNe_KS_AD}).
\item When considering early- and late-type spiral galaxies separately, the vertical distributions of
      Type Ia and CC SNe are consistent with both the ${\rm sech}^{2}$ and ${\rm exp}$ profiles
      (Table~\ref{tableallSNe}).
      In wider morphological bins (S0--Sd or Sa--Sd), the vertical distribution of Type Ia SNe
      is not consistent with ${\rm sech}^{2}$ profile,
      most probably due to the earlier and wider morphological distribution of SNe Ia host galaxies
      in comparison with CC SNe hosts (Table~\ref{table_SN_morph}),
      and the systematically thinner vertical distribution of
      the host stellar population from early- to late-type discs.
\item By narrowing the host morphologies to the most populated Sb--Sc galaxies (close to the MW morphology)
      of our sample, we exclude the morphological biasing of host galaxies between the SN types and
      the dependence of scale height of host stellar population on the morphological type.
      In these galaxies, we find that the ${\rm sech}^{2}$ scale heights
      ($\tilde{z}_{0}^{\rm SN}$) of Type Ia and CC SNe are
      $0.096\pm0.016$ and $0.042\pm0.007$, respectively.
      The ${\rm exp}$ scale heights ($\tilde{H}_{\rm SN}$) are
      $0.065\pm0.012$ and $0.028\pm0.003$, respectively.
      In Sb--Sc galaxies, the vertical distribution of CC SNe is significantly different from that
      of Type Ia SNe (Table~\ref{diffSNe_KS_AD}), being $2.3\pm0.5$ times more concentrated
      to the plane of the host disc (Table~\ref{tableallSNe}).
\item In Sb--Sc hosts, the ${\rm exp}$ scale height
      (also the $h_{\rm SN}/H_{\rm SN}$ ratio) of CC SNe is consistent with that of the
      younger stellar population in the thin disc of the MW, derived from star counts,
      while the scale height (also the ratio) of SNe Ia is consistent with that of the
      old population in the thick disc of the MW (Tables~\ref{scaleheightsvsothers} and \ref{hhzvsothers}).
\item For the first time, we show that the ratio of scale lengths to scale heights
      ($h_{\rm SN}/z_{0}^{\rm SN}$) of the distribution of CC SNe is consistent
      with those of the resolved young stars with
      ages from $\sim10$~Myr up to $\sim100$~Myr in nearby edge-on galaxies and
      the unresolved stellar population of extragalactic thin discs (Table~\ref{hz0vsothers}).
      On the other hand, the corresponding ratio for Type Ia SNe is consistent and located between
      the values of the same ratios of the two populations of resolved stars
      with ages from a few 100~Myr up to a few Gyr and from a few Gyr up to $\sim10$~Gyr,
      as well as with the unresolved population of the thick disc of nearby edge-on galaxies.
\end{enumerate}

All these results can be explained considering the age-scale height relation of
the distribution of stellar population and
the mean age difference between Type Ia and CC SNe progenitors.

\section*{Acknowledgements}

We would like to thank the anonymous referee for his/her commentary,
and also Massimo Della Valle for his constructive
comments on the earlier drafts of this manuscript.
AAH, LVB, and AGK acknowledge the hospitality of the
Institut d'Astrophysique de Paris (France) during their
stay as visiting scientists supported by
the Programme Visiteurs Ext\'{e}rieurs (PVE).
This work was supported by the RA MES State Committee of Science,
in the frames of the research project number 15T--1C129.
AAH is also partially supported by the ICTP.
VA acknowledges the support from
Funda\c{c}\~ao para a Ci\^encia e Tecnologia (FCT)
through national funds and from FEDER through COMPETE2020
by the following grants UID/FIS/04434/2013 \& POCI-01-0145-FEDER-007672,
and the support from FCT through Investigador FCT contract
IF/00650/2015/CP1273/CT0001.
This work was made possible in part by a research grant from the
Armenian National Science and Education Fund (ANSEF)
based in New York, USA.
Funding for SDSS-III has been provided by the Alfred P.~Sloan Foundation,
the Participating Institutions, the National Science Foundation,
and the US Department of Energy Office of Science.
The SDSS--III web site is \href{http://www.sdss3.org/}{http://www.sdss3.org/}.
SDSS--III is managed by the Astrophysical Research Consortium for the
Participating Institutions of the SDSS--III Collaboration including the
University of Arizona, the Brazilian Participation Group,
Brookhaven National Laboratory, University of Cambridge,
University of Florida, the French Participation Group,
the German Participation Group, the Instituto de Astrofisica de Canarias,
the Michigan State/Notre Dame/JINA Participation Group,
Johns Hopkins University, Lawrence Berkeley National Laboratory,
Max Planck Institute for Astrophysics, New Mexico State University,
New York University, Ohio State University, Pennsylvania State University,
University of Portsmouth, Princeton University, the Spanish Participation Group,
University of Tokyo, University of Utah, Vanderbilt University,
University of Virginia, University of Washington, and Yale University.

\bibliography{snbibV}

\section*{Supporting information}

Additional Supporting Information may be found in
the online version of this article:\\
\\
\textbf{PaperVonlinedata.csv}
\\
\\
Please note: Oxford University Press is not responsible for the
content or functionality of any supporting materials supplied by
the authors. Any queries (other than missing material) should be
directed to the corresponding author for the article.

\label{lastpage}

\end{document}